\documentclass{article}
\usepackage[T1]{fontenc}
\usepackage{spconf,amsmath,graphicx,booktabs,url}
\usepackage{pgfplots}
\usepackage{float}
\usepackage{algorithm,algpseudocode}
\usepackage{balance}
\usepackage{colortbl}
\usepackage{eso-pic}
\AddToShipoutPictureFG*{%
  \put(0,34){\makebox[\paperwidth]{\parbox{6.4in}{\centering\footnotesize
    This work has been submitted to the IEEE ICASSP for possible publication.
    Copyright may be transferred without notice, after which this version
    may no longer be accessible.}}}}
\algrenewcommand\algorithmicindent{1em}
\algrenewcommand\alglinenumber[1]{\normalfont\normalsize #1:}
\usetikzlibrary{arrows.meta,positioning}
\definecolor{asrblue}{HTML}{0072B2}
\definecolor{spkorange}{HTML}{D55E00}
\definecolor{jointgreen}{HTML}{009E73}
\pgfplotsset{compat=1.18}
\newcommand{\method}{RAWD-TTS}
\newcommand{\ablationcheckedrows}{%
1 & 4 & 8 & 2.82 & 0.736\\
3 & 4 & 8 & 2.55 & 0.733\\
\rowcolor{black!8} 5 & 4 & 8 & \textbf{2.47} & \textbf{0.738}\\
\midrule
3 & 8 & 4 & 2.72 & 0.733\\
3 & 2 & 16 & 2.59 & 0.725\\
}

\newcommand{\tableoneextendedrows}{%
CosyVoice3 & 5.34 & 0.738 & 7.87 & \textbf{0.770} & 3.41\\
Qwen3-TTS & 4.69 & 0.734 & 6.18 & 0.701 & \textbf{3.67}\\
\midrule
OmniVoice & 3.18 & 0.733 & 4.38 & 0.704 & 3.18\\
OmniVoice+SFT & 2.94 & \textbf{0.765} & 4.11 & 0.726 & 3.20\\
\midrule
ASR-only & \textbf{2.35} & 0.705 & \textbf{3.72} & 0.688 & 3.29\\
Speaker-only & 3.34 & 0.763 & 4.52 & 0.734 & 3.21\\
Joint $K=3$ (selected) & 2.41 & 0.736 & 3.90 & 0.713 & 3.24\\
Joint $K=3$ (final) & 2.82 & 0.743 & -- & -- & --\\
Joint (wd1) $K=3$ & 2.61 & 0.732 & 3.89 & 0.708 & 3.35\\
\rowcolor{black!8} Joint $K=5$ (selected) & 2.42 & 0.748 & 3.76 & 0.724 & 3.25\\
Joint $K=5$ (final) & 2.58 & 0.755 & 3.97 & 0.727 & 3.22\\
}

\title{RAWD-TTS: RATIO-FREE REWARD ALIGNMENT FOR\\DISCRETE-DIFFUSION VOICE CLONING}
\name{Maxim Maslov$^{1}$ \quad Kirill Borodin$^{1,2}$ \quad Vasilii Kudryavtsev$^{1,2}$ \quad Nikita Vasiliev$^{1,2}$ \quad Grach Mkrtchian$^{2,3}$}
\address{$^{1}$lab260, Yerevan, Armenia \qquad $^{2}$BitmanagerAI, Dubai, UAE \qquad $^{3}$MTUCI, Moscow, Russia\\
kborodin.research@gmail.com}

\begin{document}
\ninept
\renewcommand{\footnotesize}{\fontsize{9}{10.5}\selectfont}
\maketitle
% Keep natural heading/paragraph spacing when a table moves to the next column.
\raggedbottom

\begin{abstract}
Zero-shot text-to-speech synthesizes new utterances in a speaker's voice
from a short reference recording. Voice cloning requires accurate content
and preserved speaker identity, but supervised acoustic-token prediction
does not directly optimize these waveform-level properties. Reward-based
post-training addresses this mismatch, but in discrete diffusion, token
choices and reveal positions jointly define the sampling trajectory,
complicating alignment. We introduce \method{} (Ratio-free
Advantage-Weighted Denoising), which scores decoded samples with recognition
and speaker rewards and uses group-relative advantages to weight masked-token
reconstruction of those samples, without reverse-trajectory likelihoods or
target audio. On 500 Russian CV3-Eval voice-cloning prompts, joint alignment
reduces word error rate from 3.18\% to 2.42\% at the reward-selected
checkpoint (24.0\% relative) and to 2.58\% at the final checkpoint (19.0\%),
while WavLM speaker cosine rises from 0.733 to 0.748 and 0.755. Controlled
experiments characterize recognition--identity trade-offs and the effects of
corruption count, group composition, and weighting.
\end{abstract}


\section{Introduction}
\label{sec:intro}

Zero-shot text-to-speech (TTS) must reproduce both the requested words and the
identity of an unseen reference speaker. Supervised token prediction provides
indirect supervision for these goals: a low acoustic-token loss does not
directly measure intelligibility or speaker preservation in a decoded waveform.
Reward-based post-training addresses this mismatch by evaluating generated
speech with task-specific models. Recent flow-matching TTS alignment methods
optimize recognition and speaker rewards through denoising trajectories
\cite{sun2025f5r,wang2026flowtts}.

Discrete-diffusion TTS introduces a different optimization interface.
OmniVoice generates multiple acoustic codebooks through iterative masked
prediction, conditioned on target text and reference speech
\cite{zhu2026omnivoice}. Token choices and the order of revealing positions
jointly determine a decoding path. A final waveform therefore cannot simply be
assigned the product of autoregressive token probabilities. Trajectory-based
diffusion policy optimization can explicitly model transitions
\cite{black2024ddpo}, but requires an appropriate likelihood for the actual
sampling process.

\method{} (Ratio-free Advantage-Weighted Denoising)\footnote{Training code and configuration are available at \url{https://github.com/lab260ru/RAWD-TTS}.} uses final generated samples as reward-weighted supervision.
For each prompt, it reconstructs forward-masked acoustic tokens with
group-relative weights: positive for above-average samples and negative for
below-average samples. This decouples learning from the reverse trajectory.
Training needs only reference speech and target text, not target audio.

Building on ratio-free weighted diffusion optimization, we make three
contributions: (i) an adaptation of signed final-sample denoising to
conditional multi-codebook TTS that preserves voice-cloning conditioning and
normalizes reconstruction across codebooks and utterances; (ii) a controlled
study on a fixed backbone of reward composition, corruption count, group
allocation, and the weighting rule, scored by both training and independent
evaluators; and (iii) released training code.

\begin{figure}[H]
\centering
\begin{tikzpicture}[
  font=\small,
  box/.style={draw=black!45,rounded corners=2pt,align=center,inner sep=4pt,minimum height=7mm},
  flow/.style={-{Stealth[length=1.5mm]},semithick,draw=black!65},
  learn/.style={-{Stealth[length=1.5mm]},thick,draw=jointgreen}]
\node[box,fill=black!3,text width=7.0cm] (cond) at (0,0)
  {Target text + reference speech and transcript};
\node[box,fill=asrblue!9,text width=7.0cm] (sample) at (0,-1.1)
  {Stochastic policy $p_\theta$\\Final tokens $\{x_0^i\}_{i=1}^G$};
\node[box,fill=spkorange!9,text width=3.15cm] (reward) at (-1.95,-2.55)
  {Decode speech\\ASR + speaker rewards};
\node[box,fill=jointgreen!9,text width=3.15cm] (mask) at (1.95,-2.55)
  {$K$ forward masks\\Token reconstruction};
\node[box,fill=spkorange!9,text width=3.15cm] (adv) at (-1.95,-3.85)
  {Group-relative\\advantages $A$};
\node[box,fill=jointgreen!14,text width=3.15cm] (loss) at (1.95,-3.85)
  {Weighted loss $A\ell_\theta$\\Update $p_\theta$};
\draw[flow] (cond) -- (sample);
\draw[flow] (sample.south) -- ++(0,-0.22) -| (reward.north);
\draw[flow] (sample.south) -- ++(0,-0.22) -| (mask.north);
\draw[flow] (reward) -- (adv);
\draw[flow] (adv) -- (loss);
\draw[learn] (mask) -- (loss);
\end{tikzpicture}
\caption{\method{} connects final-sample rewards to masked-token learning.
Generated tokens provide both waveform rewards and reconstruction targets.
Advantages weight reconstruction losses; only this path updates the policy.}
\label{fig:method}
\end{figure}
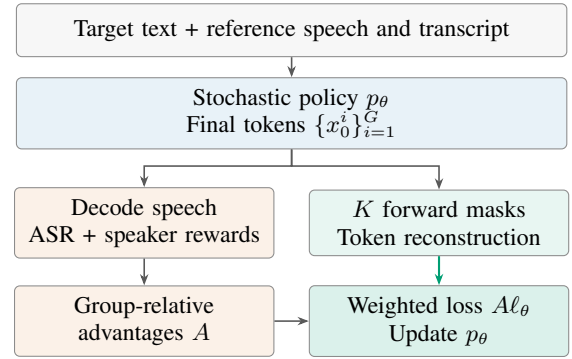

\section{Related Work}
\label{sec:related}

\textbf{Reward alignment for speech.}
F5R-TTS applies group-relative policy optimization to flow-matching speech
synthesis \cite{sun2025f5r}. Multi-objective waveform rewards are studied in
FlowTTS-GRPO \cite{wang2026flowtts}, and Koel-TTS aligns an autoregressive
TTS model with recognition and speaker-verification preferences
\cite{hussain2025koel}. These works motivate our comparison of recognition
and speaker rewards, but continuous flow and autoregressive transitions differ
from categorical masked-token updates. Masked generative codec models such as
MaskGCT \cite{wang2024maskgct} and OmniVoice \cite{zhu2026omnivoice} share
this interface.

\textbf{Weighted diffusion training.}
Ratio-free weighted denoising in wd1 \cite{tang2026wd1} and Advantage Weighted
Matching \cite{xue2026awm} are close methodological precedents. UDM-GRPO also
constructs forward-corrupted states from final samples, while retaining a
ratio-based objective \cite{wang2026udmgrpo}. SEPO derives policy gradients
for discrete diffusion from score entropy \cite{zekri2025sepo}, d1 applies
GRPO to masked diffusion language models with an approximate sequence
likelihood \cite{zhao2025d1}, and MaskGRPO estimates importance weights for
discrete diffusion \cite{ma2025maskgrpo}. The weighting principle itself
follows wd1 and AWM; our contribution is its adaptation to multi-codebook
speech with waveform rewards and the resulting empirical study.

\section{Method}
\label{sec:method}

\subsection{Voice-cloning rollouts and terminal rewards}

A prompt $q=(y,a^{\rm ref},y^{\rm ref})$ consists of target text, reference
audio, and its transcript. The native voice-cloning sampler generates acoustic
tokens $x_0\in\mathcal V^{C\times L}$ for $C$ codebooks and $L$ target frames;
codec decoding gives waveform $a=D(x_0)$. The reference and text conditioning
are fixed within a group, and target duration follows the native estimator.

Each iteration samples $B$ prompts and $G$ independent stochastic rollouts per
prompt from the current, pre-update model. Let $a_{bi}$ denote rollout $i$ for
prompt $b$. A frozen recognizer produces transcript $\hat y_{bi}$, and a frozen
speaker encoder $e$ embeds the generated and reference audio. We define
\begin{align}
r^{\rm asr}_{bi} &= \exp\!\left[-2.5\,\mathrm{WER}
 (N(y_b),N(\hat y_{bi}))\right],\label{eq:asr}\\
r^{\rm spk}_{bi} &= \frac{1+\cos(e(a_{bi}),e(a_b^{\rm ref}))}{2},\label{eq:spk}\\
r_{bi} &= \lambda_{\rm asr}r^{\rm asr}_{bi}
          +\lambda_{\rm spk}r^{\rm spk}_{bi},\label{eq:reward}
\end{align}
where $N$ applies the same Russian normalization to both texts and the
coefficient 2.5 is an empirical choice. Each component
is in $[0,1]$. Joint training uses $(\lambda_{\rm asr},\lambda_{\rm spk})=(1,1)$,
so its reward is in $[0,2]$; single-reward runs set one coefficient to zero.
We do not standardize reward components separately.

For each prompt, we compute detached advantages
\begin{equation}
\bar r_b=\frac{1}{G}\sum_i r_{bi},\qquad
A_{bi}=\frac{r_{bi}-\bar r_b}
{\sqrt{G^{-1}\sum_i(r_{bi}-\bar r_b)^2}+\epsilon},
\label{eq:advantage}
\end{equation}
with $\epsilon=10^{-6}$ and population standard deviation. Thus
$\sum_i A_{bi}=0$: comparisons are prompt-local, while standardization
also rescales each group's contribution. Constant-reward
groups contribute zero to the surrogate gradient.

\subsection{Final-sample denoising surrogate}
\label{sec:surrogate}

We detach the final tokens and draw $K$ forward corruptions for each rollout.
For state $k$, one ratio $\rho_{bk}\sim\mathcal U(0,1)$ is shared within prompt
group $b$; target-token masks are independent Bernoulli draws with that ratio.
The shared ratio acts as common random numbers: rollouts of one prompt are
compared at the same corruption level, so advantage-weighted differences
reflect rewards rather than mask-ratio variation.
Text, reference tokens, and padding are never masked. Empty target masks are
redrawn up to eight times at the same ratio; if still empty, one uniformly
chosen target token is masked. This nonempty safeguard slightly modifies the
unconditional Bernoulli corruption law.

Let $\tilde x_{bik}$ be the corrupted input, $M_{bikc}$ its masked positions in
codebook $c$, and $\mathcal C_{bik}$ the codebooks with nonempty masks. With
the base model's codebook weights renormalized over $\mathcal C_{bik}$, denoted
$\tilde w_{bikc}$, the reconstruction loss is
\begin{equation}
\ell_{bik}(\theta)=
-\sum_{c\in\mathcal C_{bik}}\frac{\tilde w_{bikc}}{|M_{bikc}|}
\sum_{j\in M_{bikc}}\log p_\theta(x_{0,bcj}^{i}\mid\tilde x_{bik},q_b).
\label{eq:denoise}
\end{equation}
This reduction first averages masked tokens within each codebook, then
combines codebooks. Longer utterances therefore do not receive larger weights
merely because they contain more tokens. The update minimizes
\begin{equation}
\mathcal L(\theta)=\frac{1}{BGK}
\sum_{b=1}^{B}\sum_{i=1}^{G}\sum_{k=1}^{K}
\operatorname{stopgrad}(A_{bi})\,\ell_{bik}(\theta).
\label{eq:objective}
\end{equation}
Gradients accumulate over all groups and corruption states before a single
optimizer step, followed by fresh rollout collection. We use neither KL
regularization nor a supervised anchor loss.

\begin{algorithm}[H]
\caption{One \method{} update}
\label{alg:update}
\begin{algorithmic}[1]
\Require Current policy $p_\theta$; batch dimensions $B,G,K$
\Statex \textit{Collect samples and scores without gradients}
\State Sample $B$ prompts $q_b$.
\State Draw $G$ native rollouts $x^i_{0,b}$ per prompt.
\State Decode $a_{bi}=D(x^i_{0,b})$ with the frozen codec.
\State Compute $r_{bi}$ using frozen scorers, Eq.~\eqref{eq:reward}.
\State Compute group advantages $A_{bi}$, Eq.~\eqref{eq:advantage}.
\Statex \textit{Differentiate masked-token reconstruction}
\State Initialize accumulated gradient $g\gets 0$.
\For{$k=1,\ldots,K$}
  \State Sample group ratios $\rho_{bk}\sim\mathcal U(0,1)$.
  \State Corrupt detached targets as in Section~\ref{sec:surrogate}.
  \State Compute $\ell_{bik}(\theta)$ using Eq.~\eqref{eq:denoise}.
  \State $g\gets g+\frac{1}{BGK}\sum_{b,i}A_{bi}\nabla_\theta\ell_{bik}(\theta)$.
\EndFor
\State Clip $g$; update $\theta$ with one AdamW step.
\end{algorithmic}
\end{algorithm}

\subsection{Interpretation and weighting control}

Equation~\eqref{eq:objective} transfers relative reward through reconstruction
of final samples. Because $\rho_{bk}$ is uniform on $(0,1)$ and
$|M_{bikc}|\approx\rho_{bk}L$, each term in Eq.~\eqref{eq:denoise} is, up to
the codebook weights and the nonempty safeguard, a length-normalized
single-sample estimate of the masked-diffusion evidence lower bound (ELBO)
with the $1/t$ weighting of MDLM, MD4, and LLaDA
\cite{sahoo2024mdlm,shi2024md4,nie2025llada}. The objective is therefore a
group-baselined REINFORCE surrogate in which a variational bound on
$\log p_\theta(x_0\mid q)$ replaces the intractable likelihood of the native
sampler. The parameter $K$ controls Monte Carlo estimation over
forward corruptions, independently of the number of reverse decoding steps.
Because a bound replaces the trajectory likelihood, the
objective is a signed surrogate rather than an unbiased policy-gradient
estimator. It imposes no likelihood-ratio constraint on native position
selection.

Our wd1-style control uses the alternative coefficients
\begin{equation}
\omega_{bi}=\frac{e^{A_{bi}}}{\sum_j e^{A_{bj}}}
             -\frac{e^{-A_{bi}}}{\sum_j e^{-A_{bj}}}.
\label{eq:wd1}
\end{equation}
This wd1-inspired weighting control \cite{tang2026wd1} retains the reduction
in Eq.~\eqref{eq:objective} without rescaling $\omega$, changing both the
relative coefficients and their gradient scale.

\section{Experimental Setup}
\label{sec:setup}

\subsection{Training data and implementation}

We use 8,000 Russian voice-cloning prompts from the YouTube partition
processed by Balalaika \cite{borodin2026balalaika}.\footnote{\raggedright\url{https://huggingface.co/datasets/lab260/youtube_balalaika}}
References are filtered for single-speaker segments
lasting 2--8~s with ASR consistency scores of at least 80. They comprise
% TODO (authors): state the scale of the ASR consistency score (e.g., 0--100).
10.96~h of audio, with mean duration 4.93~s. Each reference audio/transcript
is paired with target text from a different eligible record, restricted to
20--120 characters. Target-record audio is not used for training or rewards.

All trained systems initialize from base OmniVoice (0.6B), which predicts
\mbox{$C=8$} Higgs-audio codebooks per frame \cite{zhu2026omnivoice}, using LoRA
\cite{hu2022lora} with rank 16, alpha 32, and zero dropout; the codec is frozen.
% TODO (authors): name the LoRA target modules and the codec frame rate here.
% The OmniVoice paper states 8-codebook Higgs-audio tokenizer v2 but not the
% frame rate; the repository was not reachable to check target_modules.
Controlled reward and weighting comparisons use \mbox{$B=4$}, \mbox{$G=8$}, and \mbox{$K=3$}.
Joint \mbox{$K=5$} retains these training settings except for the corruption count. AdamW uses learning rate
$10^{-5}$ after 10 linear warmup updates, zero weight decay, and gradient-norm
clipping at 1. Training rollouts use 32 denoising steps, no classifier-free
guidance (CFG), token temperature 1, and position temperature 5, which is
applied to token confidences when sampling the positions to reveal
\cite{zhu2026omnivoice}.

On one NVIDIA RTX 6000 Ada Generation GPU with co-located reward models,
the \mbox{$K=5$} run took 9.25~GPU-hours for 2,000 training iterations
(16.65~s/update), excluding evaluations.

\subsection{Evaluation protocol}

Evaluation uses all 500 prompts of the Russian subset of
\textit{CV3-Eval Multilingual Voice Cloning}, introduced with CosyVoice~3
\cite{du2025cosyvoice3}. OmniVoice variants use the native
reference-audio/transcript prompt path, native duration estimation, 32 decoding
steps, CFG~2, greedy token selection, position temperature~5, and decoding seed
12345. Metrics are mean per-utterance WER from faster-whisper with
Whisper-large-v3 \cite{radford2022whisper}, and mean cosine similarity between
WavLM-ECAPA embeddings \cite{chen2022wavlm,desplanques2020ecapa} of generated
and reference audio; the training reward applies the mapping of
Eq.~\eqref{eq:spk} to the same cosine. WER is computed after the same
normalization of reference text and ASR predictions and is reported as a
percentage.
These scorers also supply training rewards. For an independent assessment,
Table~\ref{tab:main} additionally reports WER from Qwen3-ASR-1.7B
\cite{qwen2026asr}, raw speaker cosine from ERes2Net
\cite{chen23o_interspeech}, and predicted quality from DistilMOS
\cite{yang2026distilmos} with its WavLM backbone (raw predicted MOS, not a
listening-test score); none participates in training or checkpoint
selection. Qwen uses greedy decoding with Russian specified and no
transcript hint; all three process 16-kHz audio. Two non-terminating Qwen3-TTS outputs receive WER~$=1$;
outputs over 60~s are scored in 30-s chunks with concatenated transcripts
and duration-weighted speaker and MOS scores.

Primary runs use 2,000 updates for recognition-only, speaker-only, and joint
rewards. Table~\ref{tab:main} reports checkpoints selected by joint
training-scorer reward on the reported prompts; final checkpoints, which
involve no selection, are labeled separately. Corruption and group-size ablations
use 1,000-update budgets. The weighting comparison and extended \mbox{$K=5$} run use
2,000 updates. Primary \mbox{$K=3$} runs are evaluated every 100 updates and \mbox{$K=5$}
every 250, so the coarser \mbox{$K=5$} grid offers fewer selection
opportunities. Ablation checkpoints maximize joint evaluation reward within
the stated budget. We conduct one training run per configuration.

The SFT control uses the same 8,000 source recordings paired with their real
transcripts, not the randomized RL target texts. It reconstructs masked real
tokens with a visible 0--30\% audio prefix, using 32 utterances per update and
the same LoRA and optimizer settings. Rewards only select its checkpoint.
Thus SFT matches source data and update budget, but receives real target audio
rather than generated supervision. Controlled alignment comparisons hold the
backbone fixed. For context, we also evaluate released
CosyVoice3-0.5B-2512 \cite{du2025cosyvoice3} and Qwen3-TTS-12Hz-1.7B-Base
\cite{hu2026qwen3tts}, with native sampling and the same reference pairs.

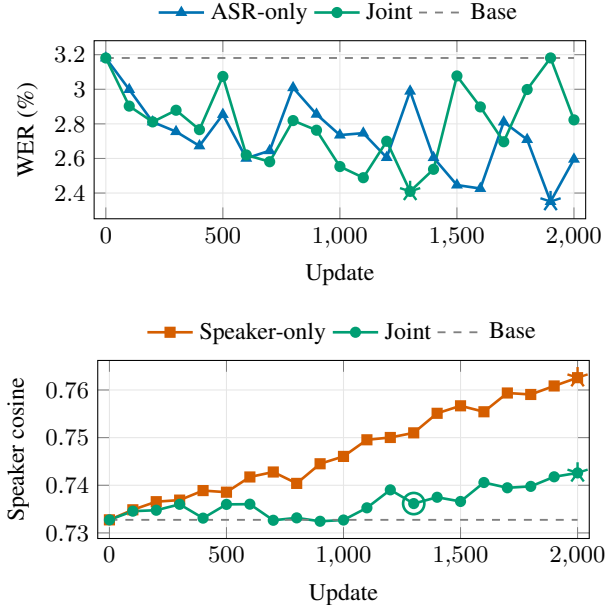
\begin{figure}[t]
\centering
\begin{tikzpicture}
\begin{axis}[width=0.94\columnwidth,height=1.55in,
xlabel={Update},ylabel={WER (\%)},
xmin=0,xmax=2000,xtick={0,500,1000,1500,2000},scaled x ticks=false,
enlarge x limits=0.025,enlarge y limits=0.12,
tick label style={font=\small},label style={font=\small},
yticklabel style={/pgf/number format/fixed,/pgf/number format/precision=3},
legend style={font=\small,draw=none,at={(0.5,1.04)},anchor=south},legend columns=3,grid=major,grid style={black!10}]
\addplot[asrblue,solid,mark=triangle*,line width=1pt,mark size=1.6pt] coordinates {(0,3.18095647) (100,2.99846748) (200,2.81107424) (300,2.75508236) (400,2.67264661) (500,2.85437506) (600,2.60084496) (700,2.64485494) (800,3.00895911) (900,2.85551041) (1000,2.73509333) (1100,2.74638717) (1200,2.60513894) (1300,2.98900688) (1400,2.60482845) (1500,2.44615310) (1600,2.42619133) (1700,2.81078241) (1800,2.70918240) (1900,2.35091447) (2000,2.59553914)};
\addlegendentry{ASR-only}
\addplot[only marks,mark=star,mark size=4pt,draw=asrblue,line width=0.9pt,forget plot] coordinates {(1900,2.35091447)};
\addplot[jointgreen,solid,mark=*,line width=1pt,mark size=1.6pt] coordinates {(0,3.18095647) (100,2.90266642) (200,2.81135396) (300,2.87879053) (400,2.76657053) (500,3.07410675) (600,2.62022768) (700,2.58105649) (800,2.81904432) (900,2.76324327) (1000,2.55320462) (1100,2.48871727) (1200,2.69936656) (1300,2.40959316) (1400,2.53729364) (1500,3.07697273) (1600,2.89803160) (1700,2.69655434) (1800,2.99862872) (1900,3.18127883) (2000,2.82308068)};
\addlegendentry{Joint}
\addplot[only marks,mark=star,mark size=4pt,draw=jointgreen,line width=0.9pt,forget plot] coordinates {(1300,2.40959316)};
\addplot[gray,dashed,line width=0.7pt] coordinates {(0,3.1809564718113084) (2000,3.1809564718113084)};
\addlegendentry{Base}
\end{axis}
\end{tikzpicture}
\par\medskip
\begin{tikzpicture}
\begin{axis}[width=0.94\columnwidth,height=1.55in,
xlabel={Update},ylabel={Speaker cosine},
xmin=0,xmax=2000,xtick={0,500,1000,1500,2000},scaled x ticks=false,
enlarge x limits=0.025,enlarge y limits=0.12,
tick label style={font=\small},label style={font=\small},
yticklabel style={/pgf/number format/fixed,/pgf/number format/precision=3},
legend style={font=\small,draw=none,at={(0.5,1.04)},anchor=south},legend columns=3,grid=major,grid style={black!10}]
\addplot[spkorange,solid,mark=square*,line width=1pt,mark size=1.6pt] coordinates {(0,0.73274376) (100,0.73485614) (200,0.73655638) (300,0.73688936) (400,0.73888586) (500,0.73854786) (600,0.74174246) (700,0.74278596) (800,0.74039356) (900,0.74451316) (1000,0.74605514) (1100,0.74954524) (1200,0.75002918) (1300,0.75100458) (1400,0.75510584) (1500,0.75667850) (1600,0.75542694) (1700,0.75936774) (1800,0.75905716) (1900,0.76084532) (2000,0.76258262)};
\addlegendentry{Speaker-only}
\addplot[only marks,mark=star,mark size=4pt,draw=spkorange,line width=0.9pt,forget plot] coordinates {(2000,0.76258262)};
\addplot[jointgreen,solid,mark=*,line width=1pt,mark size=1.6pt] coordinates {(0,0.73274376) (100,0.73456714) (200,0.73476048) (300,0.73600412) (400,0.73310310) (500,0.73599134) (600,0.73603294) (700,0.73265316) (800,0.73315756) (900,0.73244376) (1000,0.73270652) (1100,0.73526620) (1200,0.73903838) (1300,0.73616050) (1400,0.73748440) (1500,0.73660062) (1600,0.74058000) (1700,0.73947060) (1800,0.73976814) (1900,0.74177858) (2000,0.74259868)};
\addlegendentry{Joint}
\addplot[only marks,mark=star,mark size=4pt,draw=jointgreen,line width=0.9pt,forget plot] coordinates {(2000,0.74259868)};
\addplot[only marks,mark=o,mark size=4pt,draw=jointgreen,line width=1pt,forget plot] coordinates {(1300,0.73616050)};
\addplot[gray,dashed,line width=0.7pt] coordinates {(0,0.73274375) (2000,0.73274375)};
\addlegendentry{Base}
\end{axis}
\end{tikzpicture}
\caption{Evaluation WER (top) and WavLM speaker cosine (bottom) on all 500
prompts, measured every 100 updates (21 points per curve, without smoothing).
These \mbox{$K=3$} runs show each single-reward model's optimized metric and both
Joint metrics. Dashed: base model. Stars: metric optima; open circle: the
reward-selected Joint checkpoint.}
\label{fig:curves}
\end{figure}

\section{Results and Discussion}
\label{sec:results}

\subsection{Recognition and speaker-identity trade-off}

\begin{table}[!htb]
\centering
\caption{CV3-Eval Multilingual Voice Cloning (Russian, 500 prompts).
Adapters use 2,000 updates; the lower block contains RAWD-TTS
variants. Cos: raw cosine. Adapter rows not marked final are
checkpoints selected by joint training-scorer reward on these prompts
(updates 1,300 for Joint \mbox{$K=3$}, 1,500 for \mbox{$K=5$}); final rows
use update 2,000 without selection. Dashes: not scored. Bold: column optimum;
shading: highest joint training-scorer reward.}
\label{tab:main}
\small
\setlength{\tabcolsep}{3pt}
\begin{tabular}{lrrrrr}
\toprule
& \multicolumn{2}{c}{Training scorers} & \multicolumn{2}{c}{Independent} & Quality\\
\cmidrule(lr){2-3}\cmidrule(lr){4-5}
System & WER$\downarrow$ & Cos$\uparrow$ & WER$\downarrow$ & Cos$\uparrow$ & MOS$\uparrow$\\
\midrule
\tableoneextendedrows
\bottomrule
\end{tabular}
\end{table}

Within RAWD-TTS, Table~\ref{tab:main} shows that each single-reward objective improves its
target metric while degrading the complementary attribute. Recognition-only
alignment reduces WER from 3.18\% to 2.35\%, but decreases speaker cosine
from 0.733 to 0.705. Speaker-only alignment attains cosine
0.763, while WER increases to 3.34\%.

Joint \mbox{$K=3$} reduces WER to 2.41\% and increases cosine to 0.736,
preserving most of the ASR-only gain without its loss in speaker scoring.

Selected Joint \mbox{$K=5$} attains 2.42\% Whisper WER and 0.748 WavLM cosine.
It matches \mbox{$K=3$} in WER and gives slightly higher similarity. Its final
checkpoint, which involves no selection, retains most of the gain: 2.58\% WER
(19.0\% relative) and 0.755 cosine, above base on every metric in
Table~\ref{tab:main}.

\noindent\textbf{Independent assessment.}
Qwen3-ASR gives a 14.0\% relative WER reduction for Joint \mbox{$K=5$}
(4.38\% to 3.76\%), while ERes2Net cosine rises from 0.704 to 0.724.
ASR-only attains Qwen WER 3.72\% but reduces speaker cosine
to 0.688; speaker-only increases cosine to 0.734 but increases
WER to 4.52\%. Thus the recognition--identity trade-off transfers beyond
the training scorers, although the recognition gain is smaller than measured
by Whisper. All aligned variants also have DistilMOS at or above base (3.18),
so alignment does not reduce predicted quality.

\noindent\textbf{External TTS systems.}
On Russian voice cloning, base OmniVoice already attains lower recognition
error than CosyVoice3 and Qwen3-TTS under both ASR evaluators, and Joint
\mbox{$K=5$} widens this margin. CosyVoice3 attains the highest independent
speaker cosine (0.770), while Qwen3-TTS has the highest DistilMOS (3.67).
The external comparison therefore mainly reflects backbone strength on
Russian, with different recognition, identity, and quality trade-offs across
systems.

\noindent\textbf{Supervised adaptation.}
SFT attains 2.94\% Whisper WER and 0.765 WavLM cosine, compared with
2.42\% and 0.748 for Joint \mbox{$K=5$}. Independent scores give the same
trade-off: SFT has higher Qwen WER (4.11\% versus 3.76\%) and slightly
higher ERes2Net cosine (0.726 versus 0.724). Joint \mbox{$K=5$} has higher
DistilMOS (3.25 versus 3.20). SFT is therefore a strong speaker-preservation
baseline; reward alignment improves recognition, not every metric. SFT,
however, requires paired target speech, whereas reward alignment uses only
reference audio and arbitrary target text.

\noindent\textbf{Weighting rule.}
With the same joint rewards, \mbox{$K=3$}, and 2,000-update budget, wd1-style
weighting yields 2.61\% WER and 0.732 cosine, compared with 2.41\%
and 0.736 for linear signed weighting. It therefore improves neither
training metric in this comparison, but has higher DistilMOS (3.35 versus
3.24). This quality-score difference cautions against ranking models by
training rewards alone. The comparison changes both weighting shape and scale.

Figure~\ref{fig:curves} shows non-monotonic joint improvement: after the
selected update 1,300, WER rises from 2.41\% to 2.82\% at update 2,000,
while cosine improves from 0.736 to 0.743. The ASR-only run ends at 2.60\%
after its optimum of 2.35\% at update 1,900. Both training-batch rewards rise
steadily over the same run, so rising training reward does not imply
monotonic benchmark improvement.
% Figure with training-batch reward curves (training_rewards.tex) removed to
% fit the 4-page limit; restore here if space allows.

\subsection{Hyperparameter ablations}

\begin{table}[H]
\centering
\caption{Joint-reward ablations: 1,000 updates, 500 evaluation prompts,
and \mbox{$BG=32$}. The shared control uses \mbox{$K=3$}, \mbox{$B=4$}, and \mbox{$G=8$}.
Checkpoint selection and highlighting follow Table~\ref{tab:main}.}
\label{tab:ablation}
\small
\setlength{\tabcolsep}{5pt}
\begin{tabular}{lrrrr}
\toprule
$K$ & $B$ & $G$ & WER$\downarrow$ & Cos$\uparrow$\\
\midrule
\ablationcheckedrows
\bottomrule
\end{tabular}
\end{table}

\noindent\textbf{Corruption count.}
At 1,000 updates, \mbox{$K=5$} improves both metrics over \mbox{$K=1$} and \mbox{$K=3$}
(Table~\ref{tab:ablation}), motivating the extended run in Table~\ref{tab:main}.
% TODO (authors): say whether the 2,000-update K=5 run continues this run or
% is a separate run.
WER decreases with $K$, but similarity is non-monotonic. One run per setting
does not establish the reliability of these differences.

\noindent\textbf{Group composition.}
At fixed \mbox{$BG=32$} and \mbox{$K=3$}, $(B,G)=(4,8)$ gives lower WER than
$(8,4)$ or $(2,16)$. The \mbox{$G=4$} setting matches the control in
similarity, whereas \mbox{$G=16$} degrades both metrics. More alternatives per
prompt do not uniformly improve alignment: group composition trades prompt
diversity against within-prompt comparisons.

\subsection{Limitations}

The signed surrogate provides neither a likelihood-ratio trust region nor a
monotonic reward-improvement guarantee. Group standardization can amplify
small reward-model differences, and equal component coefficients do not
ensure equal gradient contributions when their within-group variations differ.
Reported checkpoints, including the SFT control, are selected by evaluation
reward on the reported prompts, which favors them relative to base and
external systems; the final-checkpoint rows are free of this selection.
Training uses stochastic rollouts without CFG, whereas evaluation uses CFG~2
and greedy tokens, so our gains show transfer to the guided evaluator rather
than direct optimization of guided generation; a guidance sweep would be
needed to separate these effects. Independent recognition and speaker models indicate
transfer of average gains, but automatic scores do not establish
human-perceived quality or freedom from reward exploitation. Evidence is
limited to one backbone, Russian speech, and one training run per
configuration. Voice cloning poses impersonation risks; releases require
appropriate consent and safeguards.

\section{Conclusion}

\method{} adapts signed final-sample denoising to discrete-diffusion voice
cloning. On CV3-Eval Russian, joint alignment with five corruption views
reduces WER by 24.0\% at the reward-selected checkpoint and by 19.0\% at the
final checkpoint, while improving speaker similarity over base OmniVoice.
Independent evaluators are consistent with both gains. Single-reward and
supervised controls reveal recognition--identity trade-offs, supporting joint
waveform rewards as a practical alignment objective for discrete TTS.

\section{Compliance with Ethical Standards}

This study used only publicly available, previously collected speech
recordings (Balalaika and CV3-Eval) and collected no new data from human
subjects, so no ethical approval was required. The authors declare no
conflicts of interest.
% TODO (authors): confirm the conflict-of-interest sentence and add a funding
% acknowledgement here if applicable (both are allowed on the fifth page).

\section{Acknowledgements}

Claude Fable 5 (Anthropic) was used to polish the text of this paper. All
ideas and experiments are the authors' own.

% \balance removed: it left the last page's right column overfull (text below
% the 229 mm text area) while the left column stayed short.
\bibliographystyle{IEEEbib}
\bibliography{references}
\end{document}